# Direct activation of PMS by highly dispersed amorphous $CoO_x$ clusters in anatase $TiO_2$ nanosheets for efficient oxidation of biomass-derived alcohols

Zhiwei Jiang[a], Zhiyue Zhao[a], Xin Li[a], Huaiguang Li[b], Hector F. Garces[c], Mahmoud Amer[d], Kai Yan[a]*

[a] *Guangdong Provincial Key Laboratory of Environmental Pollution Control and Remediation Technology, School of Environmental Science and Engineering, Sun Yat-Sen University, Guangzhou 510275, China*

[b] *School of Science and Engineering, The Chinese University of Hong Kong, Shenzhen, 518172, Shenzhen, Guangdong, China*

[c] *School of Engineering, Brown University, 182 Hope Street, Providence, USA*

[d] *Mechanical Engineering Department, Faculty of Engineering, Alexandria University, Alexandria 21544, Egypt*

* Corresponding author.

E-mail address: *yank9@mail.sysu.edu.cn.*

**Abstract:** Developing a green and cost-effective catalytic system for the selective oxidation of biomass-derived alcohols is vital for the sustainable synthesis of fine chemicals. Herein, highly dispersed subnanometric amorphous $CoO_x$ clusters in anatase $TiO_2$ nanosheets (Co-$TiO_2$) fabricated by green solvent $CO_2$-assisted approach could directly activate peroxymonosulfate (PMS) for the highly selective oxidation of various biomass-derived alcohols. Advanced characterizations (e.g., EXAFS, EPR, AC HAADF-STEM) reveal that a strong interaction of $CoO_x$ clusters and the anatase $TiO_2$ support exist in Co-$TiO_2$ and Co atom in Co-$TiO_2$ is mainly consisted of $Co^{2+}$ and $Co^{3+}$. The Co-$TiO_2$ catalyst offers superior catalytic performance in the conversion of six types of alcohols (e.g., benzyl alcohol (BAL), 5-hydroxymethylfurfural (HMF)) with high selectivity to produce corresponding aldehydes. Highly dispersed $CoO_x$ clusters and the interaction between $CoO_x$ clusters and $TiO_2$ support contribute to the superior performance. Mechanism studies show that $SO_4^{\bullet-}$ radicals play the dominant role in the selective oxidation of model reactant BAL and $^1O_2$ participates in the non-radical pathway. DFT calculations are well matched with experiment and decipher that the strong interaction between $CoO_x$ clusters and $TiO_2$ support promotes the formation of $SO_4^{\bullet-}$ and $SO_5^{\bullet-}$.

**Keywords:** $CoO_x$ clusters; Green solvent $CO_2$; Peroxymonosulfate activation; Alcohol oxidation; $TiO_2$

## 1. Introduction

The selective oxidation of alcohols to aldehydes has drawn tremendous attention since its versatile applications in organic synthesis and fine-chemical industry [1-4]. Several homogeneous reactions with stoichiometric metal oxides (such as $ClO^-$, $Cr_2O_7^{2-}$, and $MnO_4^-$) or peroxides ($H_2O_2$ and t-butyl hydroperoxide)[5-7] have been proved be an effective route for selective oxidation of alcohols. However, stoichiometric metal oxides easily result in pollution problems and peroxides suffer from low stability. By comparison, heterogeneous catalysts (such as noble metal- and non-noble metal-based catalysts) possess environmental advantages [8-10]. However, the scarcity of noble metal or low catalytic activity of non-noble metal catalysts hampered their wide application for the chemical industry [11-16].

Photocatalytic oxidation has been viewed as a promising technique because of its clean and green procedure [17, 18]. Furthermore, peroxymonosulfate (PMS, $HSO_5^-$) is a chemically stable and soluble inexpensive oxidant, which has been extensively utilized in oxidative reactions [19, 20]. Sulfate radical anions ($SO_4^{\bullet-}$) can be generated from PMS via the cleavage of the peroxide bond ($E^0(HSO_5^-/SO_4^{\bullet-}) = 2.43\ V_{NHE}$) and possess high redox potential (2.5 ~ 3.1 V) and long half-life period (30 ~ 40 μs) [21, 22]. The photocatalytic oxidation technology based on PMS activation have been extensively utilized in pollutants degradation and alcohols oxidation due to the generation of reactive $SO_4^{\bullet-}$ radicals, which show excellent activity in advanced oxidation processes [23-26]. Based on the high input energy from ultraviolet or heat to activate PMS, the development of supported transition metals (Cu, Fe, Co) with multiple valence states has received growing attention [27-29]. It is worth noting that Co-based catalysts exhibited excellent activation of PMS through the redox cycle of Co to produce highly reactive $SO_4^{\bullet-}$. However, these Co-based catalysts suffer from Co ions leaching and lead to the low catalytic stability. Resent years, $MO_x$ clusters onto oxides have been reported for catalysis because their high activity and stability [30, 31]. Although many catalytic systems for PMS activation have been reported, some of potential issues still exist: (i) PMS is indirectly activated by the extra energy input (e.g., ultraviolet or heat). (ii) The procedure of catalyst synthesis often involves several steps and is easy to induce the secondary pollution. (iii) Potential metal leaching of transitional metal under the activation process of PMS results in low stability.

Herein, highly dispersed amorphous $CoO_x$ clusters in $TiO_2$ support (Co-$TiO_2$) catalyst was successfully prepared using green solvent $CO_2$-assisted approach and these as-prepared Co-$TiO_2$ catalysts could directly activate PMS to efficiently oxidize six types of alcohols at 50 °C for 3 h. The physical properties of the Co-$TiO_2$ catalysts were studied *via* X-ray absorption fine structure (XAFS), electron paramagnetic resonance (EPR) and aberration-corrected high angle annular dark-field scanning transmission electron microscope (AC HAADF-STEM), confirming $CoO_x$ clusters strongly interacted with the anatase $TiO_2$ support. The catalytic properties of Co-$TiO_2$ catalysts were examined with various biomass-derived alcohols. For comparative purposes, four additional catalysts ($TiO_2$, $Co_3O_4$, Ni-$TiO_2$, Cu-$TiO_2$) were also evaluated. The catalytic mechanism for the selective oxidation of BAL was proposed based on DFT and in situ EPR studies.

## 2. Experimental section

### *2.1. Materials*

Tetra-n-butyl titanate ($Ti(OBu)_4$), methanol (chromatographically pure), copper nitrate trihydrate ($Cu(NO_3)_2{\cdot}3H_2O$), Cobalt (II) nitrate hexahydrate ($Co(NO_3)_2{\cdot}6H_2O$), anhydrous ethanol, Nickel (II) nitrate hexahydrate ($Ni(NO_3)_2{\cdot}6H_2O$), hydrofluoric acid (HF, 40 wt.%) were purchased from Shanghai Macklin Biochemical Co. Ltd. $CO_2$ (> 99.9%) was obtained from the Guangzhou Gas Company.

### *2.2. Preparation of Co-$TiO_2$*

Anatase $TiO_2$ nanosheets were fabricated and modified from the previously work [32]. Briefly, 10 mL of $Ti(OBu)_4$ was added into anhydrous ethanol (40 mL) containing HF (1.2 mL). After stirring for 30 min, the solution was transferred into a Teflon-lined stainless-steel autoclave with a capacity of 100 mL and then heated at 180 °C for 2 h. The white product was collected by centrifugation and subsequently washed with anhydrous ethanol and ultrapure water three times. Finally, the product was dried at 60 °C under vacuum for 12 h.

The preparation of Co-$TiO_2$ was performed as follows: the synthesized $TiO_2$ (200 mg) and a specific amount of $Co(NO_3)_2 \cdot 6H_2O$ (the atom ratio of Co to Ti is 6%) were firstly added into the glass vial and then placed in a stainless steel autoclave. Afterwards, 4 MPa $CO_2$ (>99.99%, Guangzhou Gas Company) was pumped into the autoclave and then stirred at room temperature for 24 h. The pink nanoparticles were collected through the slow release of $CO_2$ and then annealed at 500 °C in the air with a heating rate of 5 °C $min^{-1}$ for 4 h to obtain the brown nanoparticles, denoted as Co-$TiO_2$. Ni-$TiO_2$, and Cu-$TiO_2$ were fabricated through a similar procedure used for Co-$TiO_2$ except that $Co(NO_3)_2 \cdot 6H_2O$ was replaced by the corresponding $Ni(NO_3)_2 \cdot 6H_2O$ or $Cu(NO_3)_2 \cdot 3H_2O$, respectively. $Co_3O_4$ was obtained by the calcination of $Co(NO_3)_2 \cdot 6H_2O$ at 400 °C for 4 h in a muffle furnace.

*2.3. Selective oxidation of alcohols*

In a typical reaction process for the selective oxidation of benzyl alcohol, 5 mg of catalyst was added into 5 mL of acetonitrile/water solvent (1:1, volume ratio) with 20 mM of BAL in a 10 mL flask. Then, 0.12 mmol PMS was added to the flask. The solution was stirred at 700 rpm at 50 °C for 3 h. After the reaction, the flask was immediately quenched in an icy water-bath. Anisole (0.1 mmol) was injected into the flask as an internal standard. 1 mL of reaction mixture was withdrawn and the precipitate was filtered out. The substrates and products were analyzed by High-performance liquid chromatography (HPLC, Shimadzu Prominence SIL-20A) equipped with a RAD detector at a wavelength of 230 nm. Ultrapure water and methanol in the proportion of 30:70 %vol were used as mobile phase with a 0.8 m $min^-$ [1] flow rate at 40 °C. For oxidation of HMF and other alcohols, the similar reaction procedure is also employed.

*2.4. Structural characterizations*

X-ray diffraction (XRD) patterns were collected in a Rigaku Ultima IV X-ray diffractometer with Cu-Kα radiation at 40 mA and 40 kV. Inductively coupled plasma-optical emission spectrometry (ICP-OES) experiments were measured on an Agilent 5110. Transmission electron microscopy (TEM) analysis was carried out on a FEI TECNAI G2 F20 field emission transmission electron microscope with an accelerating voltage at 200 kV. Aberration corrected Transmission electron microscope (ACTEM) was conducted on JEM-ARM300F at 300 kV acceleration voltage. The XPS spectra of the samples were collected by Thermo Scientific model 250Xi spectrometer and the binding energies were referenced to the C1s line at 284.8 eV. X-band electron paramagnetic resonance (EPR) was conducted by Brooke a300. The X-ray absorption spectra of Co K-edge were collected at BL14W1 beamline of Shanghai Synchrotron Radiation Facility (SSRF) Shanghai. The data were collected in fluorescence mode using a Lytle detector while the corresponding reference sample was collected in transmission mode. XANES and EXAFS data were processed by Athena software.

*2.5. Computational methods*

All calculations were performed using the plane wave-based periodic DFT method as implemented in the Vienna Ab Initio Simulation Package (VASP) [33]. The electron-ion interaction was described with the projector augmented wave (PAW) method [34]. The electron exchange and correlation energies were treated within the generalized gradient approximation in the Perdew-Burke-Ernzerhof functional (GGA-PBE) [35]. The plane wave basis was set up to 500 eV. For the PMS molecule adsorbed on CNTs, the adsorption energy Eads is defined as

$E_{ads} = E_{PMS/Sub} – (E_{PMS} + E_{Sub})$

where $E_{PMS/Sub}$, $E_{PMS}$, and $E_{Sub}$ are the total energies of the PMS/substrate system, the isolated PMS molecule, and substrate in the same slab, respectively.

The barrier ($E_b$) and formation energy ($E_f$) were, respectively, calculated according to $E_a = G_{TS} – G_{IS}$, and $E_f = G_{FS} – G_{IS}$, where $G_{IS}$, $G_{TS}$, and $G_{FS}$ are the Gibbs free energies of the corresponding initial state (IS), transition state (TS) and final state (FS) at 300 K.

## 3. Results and discussions

### *3.1. Synthesis and characterization of Co-TiO$_2$*

Amorphous $CoO_x$ clusters dispersed in $TiO_2$ nanosheets catalyst (Co-$TiO_2$) with 6 mol% Co was synthesized using the green solvent $CO_2$-assisted method. The synthesis diagram is briefly depicted in **Fig. 1** and more details are shown in the Experimental Section. The compressed $CO_2$ was simultaneously used as a green solvent for the solution of Co salt and the dispersion medium of $TiO_2$, resulting in the uniformly dispersion of Co ions in the surface of $TiO_2$ nanosheets. The calcination of the obtained nanoparticles is essential to enhance the formation of $CoO_x$ clusters [36].

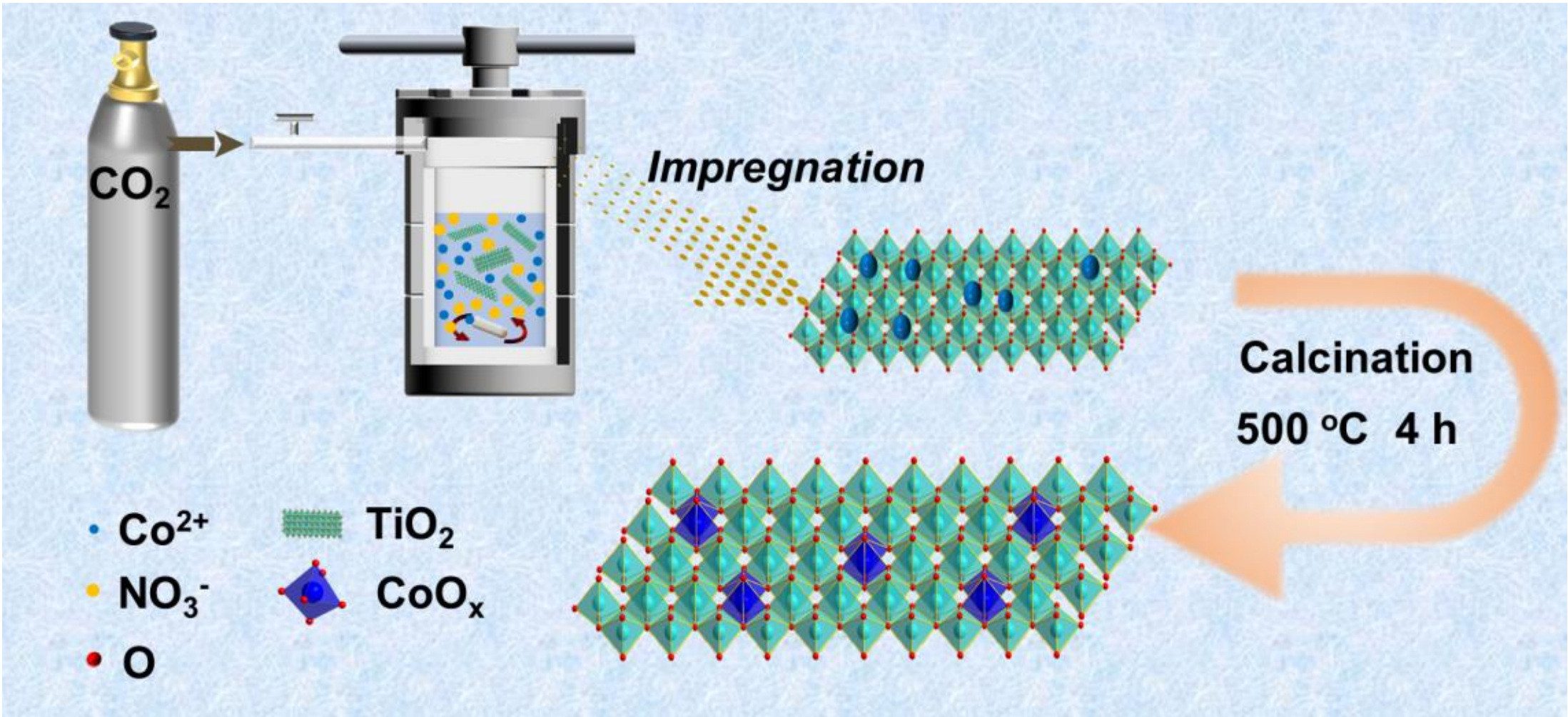

**Fig. 1.** Schematic diagram for the synthesis of Co-$TiO_2$ nanosheets.

To evaluate the textural information of Co-$TiO_2$ catalyst, transmission electron microscopy (TEM) was carried out, as shown in Fig. 2, Figs. S1 and S2. Co-$TiO_2$ catalyst exhibited a nanosheet structure with a length of 10 ~ 30 nm and a thickness of ~ 3 nm (Figs. 2a and 2b). Selected area electron diffraction (SAED) pattern further showed well-defined diffraction rings of anatase $TiO_2$ (Fig. 2c) and no visible signal of $CoO_x$ was observed, which is in accordance with the XRD results. High dispersion and small size of $CoO_x$ clusters is further revealed in Co-$TiO_2$. High-resolution TEM (HR-TEM) images further exhibited the high crystallinity of Co-$TiO_2$ with an average lattice spacing of 0.364 nm that matched with the (101) facet of anatase $TiO_2$ (Fig. 2d). The lattice spacing of $CoO_x$ is still unnoticeable. The texture of amorphous $CoO_x$ was then investigated through a high angle annular dark field combined with aberration-adjusted scanning transmission electron microscopy (HAADF-STEM) because of its sensitivity for different elements species in the composites. It turns out that amorphous $CoO_x$ clusters were highly dispersed throughout the $TiO_2$ support (white dots, Fig. 2f) in the size range of 0.7 to 1.5 nm. Meanwhile, a variation of the lattice spacing (d1: 0.364 nm, d2: 0.367 nm) in different regions indicated the appearance of lattice distortion forward (101) facet (Fig. 2e and Fig. S1) due to the strong interaction between $CoO_x$ cluster and $TiO_2$ [30]. Energy-dispersive X-ray spectroscopy (EDS) pattern showed that Ti, O, Co elements were homogeneously dispersed on the catalyst (Fig. 2g and Fig. S2a), revealing that $CoO_x$ clusters are uniformly distributed throughout the $TiO_2$ support. The AC-TEM image of Co-$TiO_2$ revealed only a crystal lattice for anatase $TiO_2$ (Fig. 2h). This suggested that highly dispersed amorphous $CoO_x$ clusters in anatase $TiO_2$ support. The atomic ratio of Co to Ti in Co-$TiO_2$ was determined as 1:9 by aberration-corrected scanning transmission electron microscopy (AC-STEM) and electron energy loss spectroscopy (EELS) analysis (Fig. S2b and Fig. 2i), close to the value of 1:8.7 obtained from the line-scan profile (Figs. S2c, d), which is larger the ratio (1:20) obtained from ICP-OES, further confirming the formation of $CoO_x$ clusters. In contrast, XPS results revealed that the atomic ratio of Co to Ti is roughly 1: 6.8 on the surface of Co-$TiO_2$ (Fig. S3), which is three times larger than that in the bulk of Co-$TiO_2$ catalyst. This discrepancy is highly possible due to the formation of amorphous $CoO_x$ clusters on the $TiO_2$ surface during high-temperature processing [30].

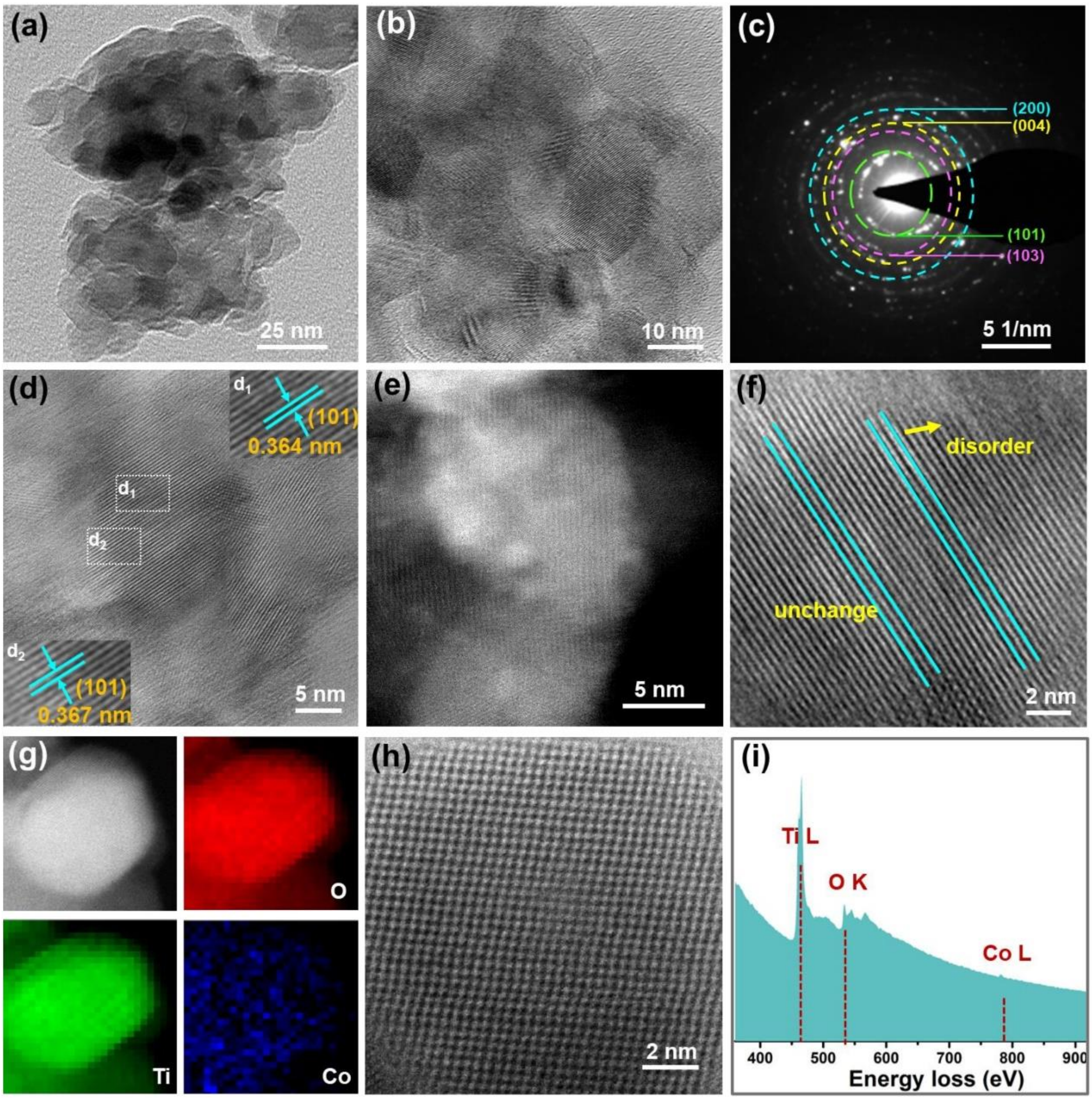

**Fig. 2.** (a, b) TEM images, (c) SAED pattern, (d) HR-TEM image, (e) HAADF-STEM image of $CoO_x$ clusters in Co-$TiO_2$. (f) lattice distortion area of Co-$TiO_2$. (g) STEM-EDS elemental maps for O, Ti, and Co, respectively. (h) AC-TEM image of Co-$TiO_2$. (i) EELS point spectrum of Co-$TiO_2$ (high-loss).

To further study the crystal structure of $CoO_x$ cluster in Co-$TiO_2$, X-ray diffraction (XRD) patterns of the fabricated $TiO_2$ and Co-$TiO_2$ were compared in Fig. S4a, where both displayed the characteristic peaks of anatase $TiO_2$ (JCPDS-21-1272). The absence of $CoO_x$ peaks in the Co-$TiO_2$ spectrum was probably due to the high dispersion of amorphous $CoO_x$ on $TiO_2$ support. To investigate the short-range surface structural change of $TiO_2$ and Co-$TiO_2$, Raman spectra were further utilised (Fig. S4b). Co-$TiO_2$ catalyst exhibited the typical anatase bands of $TiO_2$ without noticeable new peaks [31]. Compared to the $TiO_2$, the red-shifts of $E_g$ peak (144 $cm^{-1}$) and $A_{1g}$ peak (514 $cm^{-1}$) in the Co-$TiO_2$ catalyst indicate the structural change of anatase $TiO_2$ induced by the interaction between $CoO_x$ cluster with the $TiO_2$ support [30]. Furthermore, the enhanced intensity of $B_{1g}$ peak at 394 $cm^{-1}$ of Co-$TiO_2$ suggests that the interaction also influenced the symmetric bending vibration of O-Ti-O [37].

To investigate the local structure of $CoO_x$ cluster and its interaction in Co-$TiO_2$ catalyst, extended X-ray absorption fine structure (EXAFS) spectroscopy was then performed (**Fig. 3**). Fig. 3a shows the Co K-edge of the X-ray absorption near-edge structure (XANES) spectra of Co-$TiO_2$, $Co_3O_4$, Co foil and CoO. The Co K-edge of Co-$TiO_2$ is very close to $Co_3O_4$ (Inset in Fig. 3a), indicating that Co in Co-$TiO_2$ is mainly consisted of $Co^{2+}$ and $Co^{3+}$. Compared with CoO, the absorption edge of Co-$TiO_2$ is shifted higher energies, confirming that Co in $CoO_x$ clusters have a high oxidation state. Whereas the intensity of Co-$TiO_2$ is lower than that of $Co_3O_4$, indicating a different coordination environment [31]. The intensity of the shoulder peak at 7723 eV valence state for $CoO_x$ is stronger than that for $Co_3O_4$, revealing that $CoO_x$ have a relative lower electron resistance. Fourier-transformed $k^3$-weighted EXAFS spectra at Co edge in the R space is shown in Fig. 3b. The Co-$TiO_2$ curve exhibited three major peaks that are attributed to Co-O bonds (1.46 Å), Co-Co bonds (2.48 Å) and Co-M bonds (4.72 Å). The intensity of the R space curve for Co-$TiO_2$ had an evident reduction relative to $Co_3O_4$, verifying the presence of oxygen vacancies in Co-$TiO_2$. A sharp peak located at ≈ 1.0 Å for Co-$TiO_2$ was also observed due to the Co-O distortion caused by the strong interaction between $CoO_x$ cluster and $TiO_2$ support in Co-$TiO_2$. The wavelet transform (WT) contour plots are displayed in Fig. 3c (Co-$TiO_2$) and Fig. S5 ($Co_3O_4$, Co foil and CoO). Fitted EXAFS data of Co-$TiO_2$ showed that the Co-O coordination number and the Co-Co coordination number are 5.23 and 4.0 on average respectively (Fig. 3d and Table S1), confirming the existence of O vacancies.

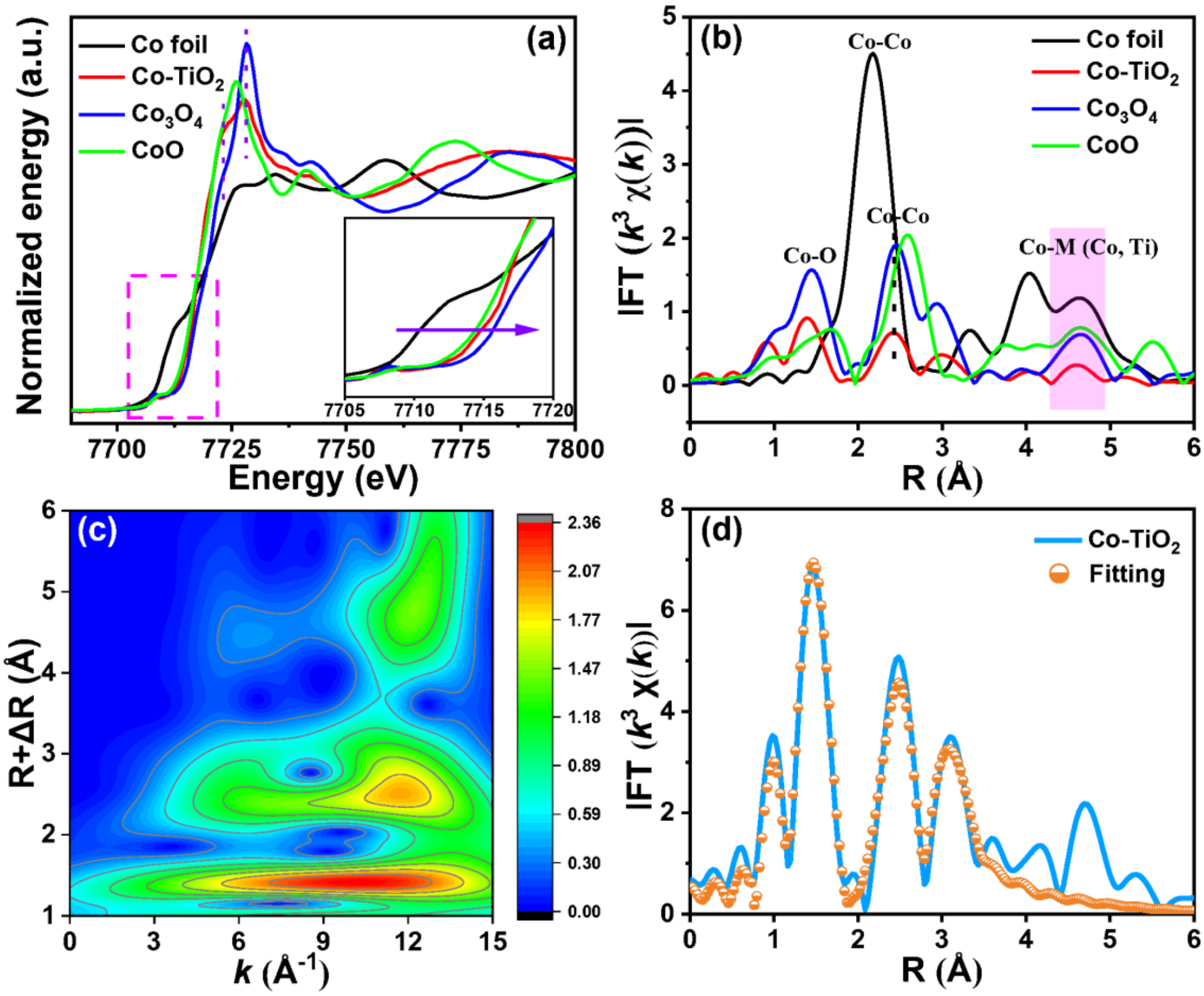

**Fig. 3.** (a) Experimental Co K-edge XANES spectra of Co-$TiO_2$ and related samples (Co foil $Co_3O_4$, and CoO). (b) Fourier-transformed magnitude of the experimental Co K-edge EXAFS signal of Co-$TiO_2$ and related samples in R space. (c) Wavelet transform (WT) for the $k^3$-weighted EXAFS signals of Co-$TiO_2$. (d) Corresponding EXAFS R-space fitting curve of Co-$TiO_2$.

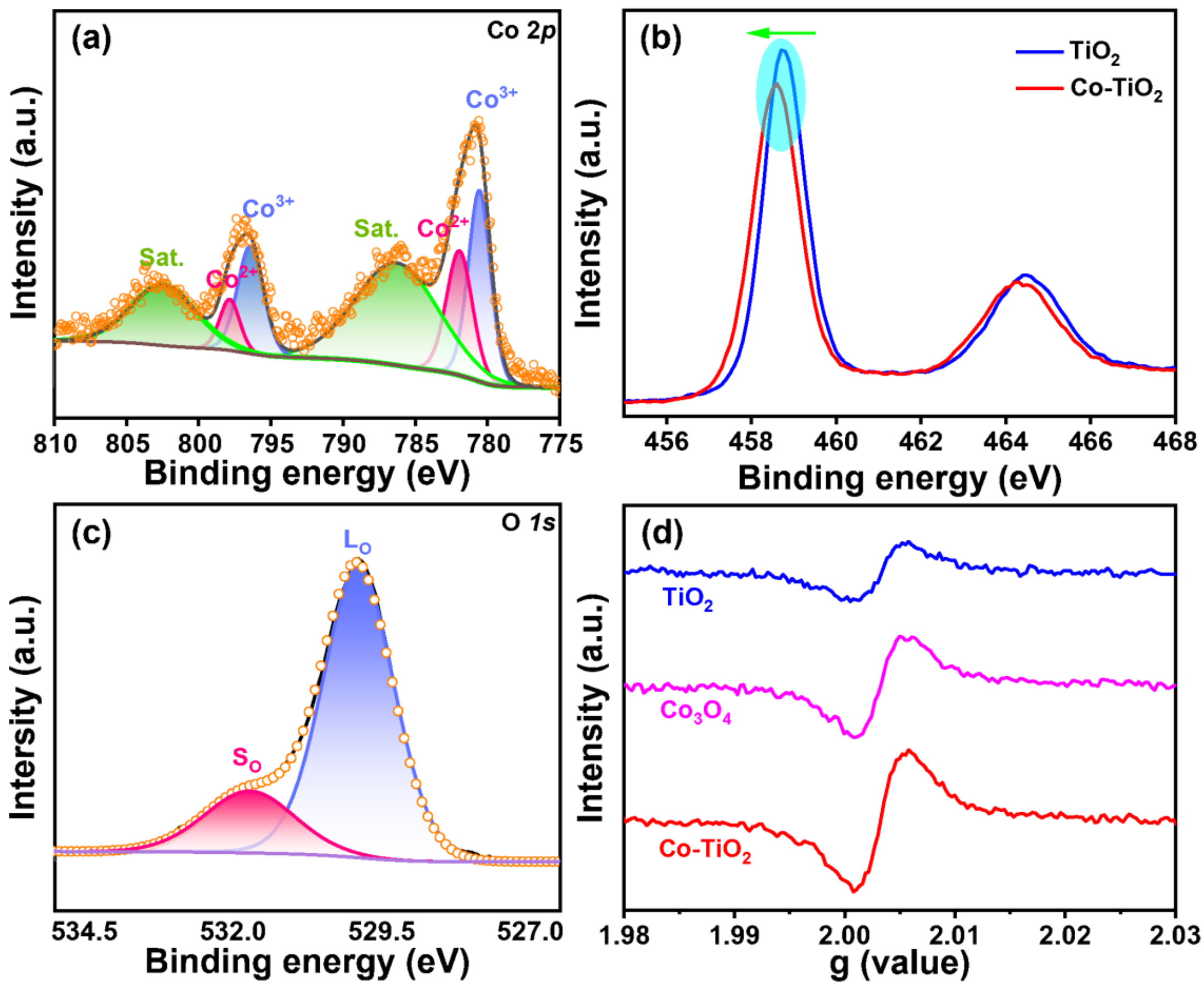

**Fig. 4.** (a) Co 2p XPS spectra of Co-$TiO_2$. (b) Ti 2p XPS spectra of Co-$TiO_2$ and $TiO_2$. (c) O 1s XPS spectra of Co-$TiO_2$. (d) EPR spectra of $Co_3O_4$, $TiO_2$, and Co-$TiO_2$.

To examined the valence states of $Co_3O_4$, $TiO_2$ and Co-$TiO_2$, XPS was applied. In Co 2p spectra of Co-$TiO_2$ (**Fig. 4**a), the peaks appeared at around 780.6 and 796.4 eV attributed to the Co $2p_{1/2}$ and Co $2p_{3/2}$ of $Co^{3+}$ in $CoO_6$ octahedron, respectively, and at around 782.5 and 798.8 eV assigned to Co $2p_{1/2}$ and Co $2p_{3/2}$ of $Co^{2+}$ in $CoO_4$ tetrahedron, respectively, consistent with the XANES result. The content of $Co^{3+}$ on the surface of Co-$TiO_2$ is larger than that of $Co^{2+}$, confirming that $CoO_6$ octahedron is the main composition for Co. For Ti 2p spectra (Fig. 4b), the peaks shift negatively from 458.8 ($TiO_2$) to 458.6 eV (Co-$TiO_2$) due to the partial reduction of $Ti^{4+}$ to $Ti^{3+}$, indicative of the Co dopant-induced surface structure change of anatase. The O 1s XPS spectra of Co-$TiO_2$, as shown in Fig. 4c, displayed two peaks at 529.9, and 531.7 eV corresponding to the lattice oxygen ($L_o$) and surface adsorbed oxygen ($S_o$), respectively. To evaluate and compare the oxygen vacancy ($V_o$) concentration in $Co_3O_4$, $TiO_2$, and Co-$TiO_2$, electron paramagnetic resonance (EPR) offered the peaks at a g-value of 2.003, attributing to $V_o$, in the three samples (Fig. 4d). The highest peak intensity in Co-$TiO_2$ catalyst confirmed that the Co dopant enhanced the $V_o$ content of the $TiO_2$ support. The presence of $V_o$ and $Ti^{3+}$ further improved the charge carrier density over Co-$TiO_2$, which is beneficial for the PMS activation [21].

*3.2. Selective oxidation of alcohols*

The selective oxidation of BAL to BZH was firstly studied to evaluate the catalytic properties of a range of catalysts, at 50 °C in the presence of PMS (**Fig. 5**). The catalyst-free system with only PMS afforded less than 4% conversion. Besides, $TiO_2$ support showed very limited BZH selectivity (29.7%) with 6.4% conversion of BAL. While, the BZH selectivity increased to 84.2% after using $Co_3O_4$ particles, but the reaction is still modest-yielding. A largely promoted BAL conversion (over 52%) with high selectivity (91.1%) toward BZH was achieved using Co-$TiO_2$ as catalyst, compared to the counterpart $Co_3O_4$ particles. This result revealed that the synergetic interaction between $CoO_x$ and $TiO_2$ support with a high concentration of $V_o$ enhanced PMS activation for the BZH conversion. In contrast, both Cu-$TiO_2$ and Ni-$TiO_2$ catalysts displayed low efficiency for BAL conversion. Besides, NiO and CuO also examined low catalytic activity. Notably, the BZH selectivity over Co-$TiO_2$ (91.1%) and $Co_3O_4$ (84.2%) was higher than that of $TiO_2$ (25%), suggesting that Co element plays mainly active site for BAL conversion and Co doping highly promoted BZH selectivity.

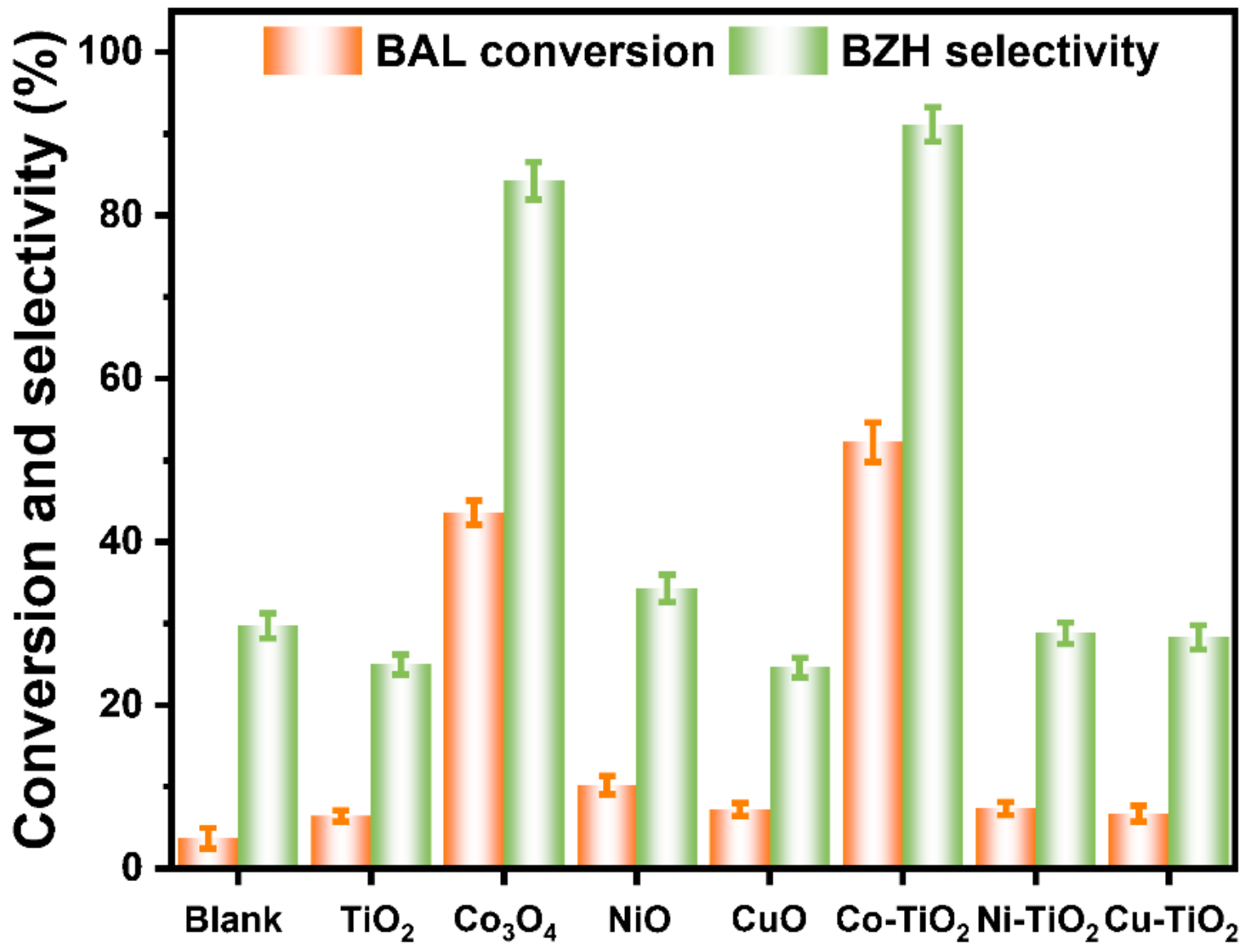

**Fig. 5.** Selective oxidation of BAL over different catalysts. Reaction conditions: 5 mg catalyst, 0.1 mmol BAL, 0.12 mmol PMS, 5 mL acetonitrile/water (1:1, volume ratio), 3 h, 50 °C.

The influence of reaction conditions in the catalytic oxidation was explored by varying reaction time, temperature, PMS dosage, and pH, as shown in Fig. S6. The BAL conversion increased from 30.6% to 60.1% within the reaction time ranging from 1 to 5 h while the BZH selectivity slightly decreased (Fig. S6a), confirming that the extension of the reaction time could improve the BZH yield. Similar trends of BAL conversion and BZH yield were exhibited with the increase of reaction temperature from 35 to 65 °C (Fig. S6b). At 65 °C, the conversion of BAL achieved 59% with a 52.2% yield of BZH. The effect of pH on the PMS activation for BAL conversion was also studied (Fig. S6c). At pH = 4, BAL conversion (36.7%) and BZH (22.1%) selectivity were much lower than that at pH = 6.8 or 9.18, suggesting that acidic conditions could restrain PMS activation and thus showed lower catalytic activity. It was noteworthy that under alkaline condition (pH = 9.18) the selectivity of BZH was increased to 93.3% with the BZH yield of 51.4%. High PMS dosage could improve greatly the BAL conversion while the BZH selectivity sharply decreased (Fig. S6d). The maximum BAL conversion with (82.9%) was observed at the PMS/BAL ratio of 3.2, while the BZH selectivity decreased from 94.5% to 60.7%, confirming that the excess PMS may promote deep oxidation of the obtained BZH. The highest BZH yield was 53.2% at the PMS/BAL ratio of 2.2. Furthermore, the reusability of Co-$TiO_2$ for BAL conversion was also investigated, as shown in Fig. S7. Co-$TiO_2$ was reused four times without notable changes, reflecting the excellent catalytic stability. In addition, XRD, Raman, and XPS experiments of the spent Co-$TiO_2$ also revealed that the structure of Co-$TiO_2$ after reaction remained basically unchanged, as shown in Fig. S8. This indicated that the Co-$TiO_2$ catalyst was very stable for the selective oxidation of BAL via PMS activation.

**Table 1.** Selective oxidation of alcohols to the corresponding aldehyde using Co-$TiO_2$ in the presence of PMS.[a]

| Substrates | Catalyst | Conversion (%) | Selectivity (%) |
|---|---|---|---|
| 2-Methyl BAL | Co-$TiO_2$ | 46.4 | 98.6 |
| 3-Methyl BAL | Co-$TiO_2$ | 45 | 86.9 |
| *p*-Methyl BAL | Co-$TiO_2$ | 48.9 | 75.5 |
| *p*-Methoxy BAL | Co-$TiO_2$ | 20.3 | 45.3 |
| HMF | Co-$TiO_2$ | 47.5 | 86.5 |

[a] Reaction conditions: 5 mg catalyst, 0.1 mmol substrate, 0.12 mmol PMS, 5 mL acetonitrile/water (1:1, volume ratio), 3 h, 50 °C.

To explore the versatility of Co-$TiO_2$ catalyst, the catalytic performance of Co-$TiO_2$ was also explored for conversion of various aromatic alcohols with different substituent positions (Table 1). For methylbenzyl alcohol with different substituent positions, the methyl group at *ortho* positions showed a higher selectivity (98.6%) to the corresponding aldehyde compared with that in the *meta* (86.9%) and *para* (75.5%) position. *p*-Methoxy BAL as substrate were oxidation under the reaction condition, no desired aldehyde was obtained, suggesting that methoxy group could restrained the PMS activation over Co-$TiO_2$ catalyst. Interestingly, 5-hydroxymethyl furan (HMF) was high selectively converted to 2,5-diformylfuran the corresponding aldehyde (86.5%). It revealed that the Co-$TiO_2$/PMS catalytic system could effectively convert various alcohols to the corresponding aldehydes.

*3.3. Mechanism study*

Radical quenching tests were conducted to confirm the reaction pathways to activate PMS for BAL oxidation and reveal the radical and non-radical contribution of BAL oxidation reactions. For the radical-based oxidation process, PMS activation routes with Co has been reported [38]. The generation routes of $^{\bullet}OH$, $SO_4^{\bullet-}$, and $SO_5^{\bullet-}$ from PMS via the electron transfer among $Co^{2+}$, $Co^{3+}$ are shown in Equations (1)-(5). $SO_4^{\bullet-}$ possesses higher redox potential (2.5~3.1 V) and a longer half-life period, which can be generated from $^{\bullet}OH$ and $SO_5^{\bullet-}$ radicals [39]. The redox cycle of $Co^{2+}/Co^{3+}$ ensured that the Co-$TiO_2$ continuously participated in the radical generation without deactivation. To verify the evolution of $^{\bullet}OH$, $SO_4^{\bullet-}$ radicals, and $^1O_2$ during the reactions, in situ electron paramagnetic resonance (EPR) was carried out using 5,5-dimethyl-1-pyrroline *N*-oxide (DMPO) and 2,2,6,6-tetramethyl-4-piperidinol (TMP) as probes to trap radicals ($^{\bullet}OH$, $SO_4^{\bullet-}$) and $^1O_2$, respectively. **Figs. 6**a and 6b display EPR spectra of $Co_3O_4$, $TiO_2$, and Co-$TiO_2$ catalyzed PMS with DMPO or TMP, finding that the peak intensities of radicals and $^1O_2$ followed the order of Co-$TiO_2$ > $Co_3O_4$ > $TiO_2$. This result demonstrated that Co-$TiO_2$ possessed an excellent activation compared to $Co_3O_4$ for PMS while it was difficult for $TiO_2$ to activate PMS, further revealing that $CoO_x$ clusters in Co-$TiO_2$ exhibited unprecedented activity for PMS.

$$Co^{2+} + HSO_5^{-} \rightarrow Co^{3+} + {}^{\bullet}OH + SO_4^{2-} \quad (1)$$

$$Co^{2+} + HSO_5^{-} \rightarrow Co^{3+} + OH^{-} + SO_4^{\bullet-} \quad (2)$$

$$Co^{3+} + HSO_5^{-} \rightarrow Co^{2+} + H^{+} + SO_5^{\bullet-} \quad (3)$$

$${}^{\bullet}OH + SO_4^{2-} \leftrightarrow OH^{-} + SO_4^{\bullet-} \quad (4)$$

$$2SO_5^{\bullet-} \rightarrow {}^1O_2 + 2SO_4^{\bullet-} \quad (5)$$

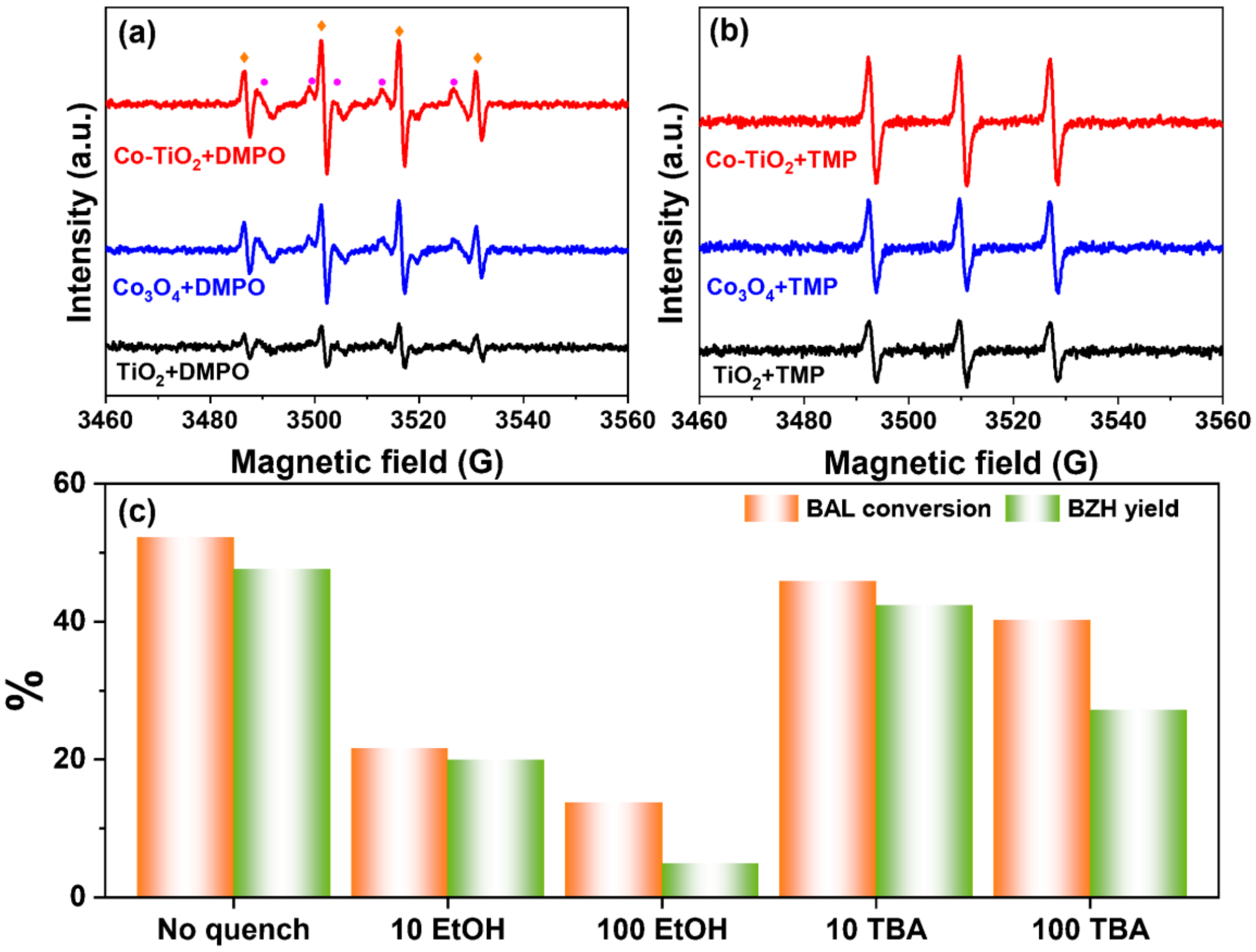

**Fig. 6.** EPR spectra of $Co_3O_4$, $TiO_2$, and Co-$TiO_2$ catalyzed PMS in the presence of (a) DMPO (DMPO-$^{\bullet}$OH—●, DMPO-$SO_4^{\bullet-}$—♦) and (b) TMP. (c) Quenching effects on the selective oxidation of BAL. Reaction conditions: 5 mg of Co-$TiO_2$, 0.1 mmol BAL, 0.12 mmol PMS, 5 mL of acetonitrile/water (1:1, volume ratio), 50 °C, 3 h.

To verify these reaction pathways, ethanol (EtOH) and *tert*-butanol (TBA) were chosen as the radical scavengers. TBA showed a higher reaction rate towards $^{\bullet}OH$ compared with $SO_4^{\bullet-}$ while EtOH could scavenge effectively both hydroxyl and sulfate radicals [40]. In Fig. 6c, the conversion of BAL over Co-$TiO_2$ decreased with a higher concentration of TBA and EtOH, but EtOH showed a stronger inhibition effect for BAL conversion than TBA, indicating that $SO_4^{\bullet-}$ radicals played a dominant effect in the BAL generation. When the scavenger dosage was 100 times of PMS, EtOH and TBA resulted in 73.7% and 23% loss of BAL conversion, respectively, roughly confirming that 23.3% out of the total BAL oxidation over Co-$TiO_2$ was contributed to the nonradical-based process. Thus, we can conclude that BAL oxidation through PMS over Co-$TiO_2$ was a radical dominated process and $SO_4^{\bullet-}$ radicals mainly contributed to BAL oxidation. Furthermore, non-radical activation of PMS included two possible pathways: singlet oxygen ($^1O_2$) and electron transfer by substrate/PMS [41]. Due to low catalytic activity of $TiO_2$ for BAL oxidation over PMS, electron transfer was not a dominant pathway in the non-radical oxidation. $^1O_2$ can be generated from Equation (5) and $^1O_2$ was detected by in situ EPR experiments over Co-$TiO_2$. Therefore, it could be inferred that $^1O_2$ participated in the non-radical oxidation of BAL.

To fully understand the roles of $Co^{2+}$, $Co^{3+}$, $Ti^{4+}$ on Co-$TiO_2$ for PMS activation, density functional theory (DFT) calculations were conducted. To simplify the model of Co-$TiO_2$, one Co atom ($Co^{3+}$) replaced a Ti atom in the anatase $TiO_2$ (101) facet and the other Co atom ($Co^{2+}$) located on the surface of the anatase $TiO_2$ (101) facet. The model of the $Co_3O_4$ (311) facet was used as a control. All possible optimum structures of PMS molecules ($HSO_5^-$) adsorbed on Co-$TiO_2$ and $Co_3O_4$ with the lowest adsorption energy ($E_{ads}$) are illustrated in **Fig. 7**. Table S2 lists the corresponding $E_{ads}$ and bond length of O-O ($l_{O\text{-}O}$) of PMS (HO-$OSO_3$) and O-H ($l_{O\text{-}H}$) of PMS (H-$OOSO_3$). The $l_{O\text{-}O}$ of PMS adsorbed on Co-$TiO_2$ and $Co_3O_4$ is larger than that of the free PMS (1.326 Å) molecules, indicating a potential formation of $^{\bullet}OH$ and $SO_4^{\bullet-}$ radicals. For Co-$TiO_2$, the maximum $E_{ads}$ and the longest $l_{O\text{-}O}$ of PMS were -1.169 eV and 1.480 Å, corresponding to the PMS-$Ti^{4+}$ in the Co-$TiO_2$ structure, implying that $Ti^{4+}$ has less activation than $Cu^{2+}$ and $Co^{3+}$ sites for PMS. Furthermore, the $E_{ads}$ and $l_{O\text{-}O}$ of PMS adsorbed on $Cu^{2+}$ and $Co^{3+}$ sites in Co-$TiO_2$ were less than that in $Co_3O_4$, revealing that small $CoO_x$ clusters doped into the surface crystal lattice of anatase $TiO_2$ have more activity for PMS than the pure $Co_3O_4$.

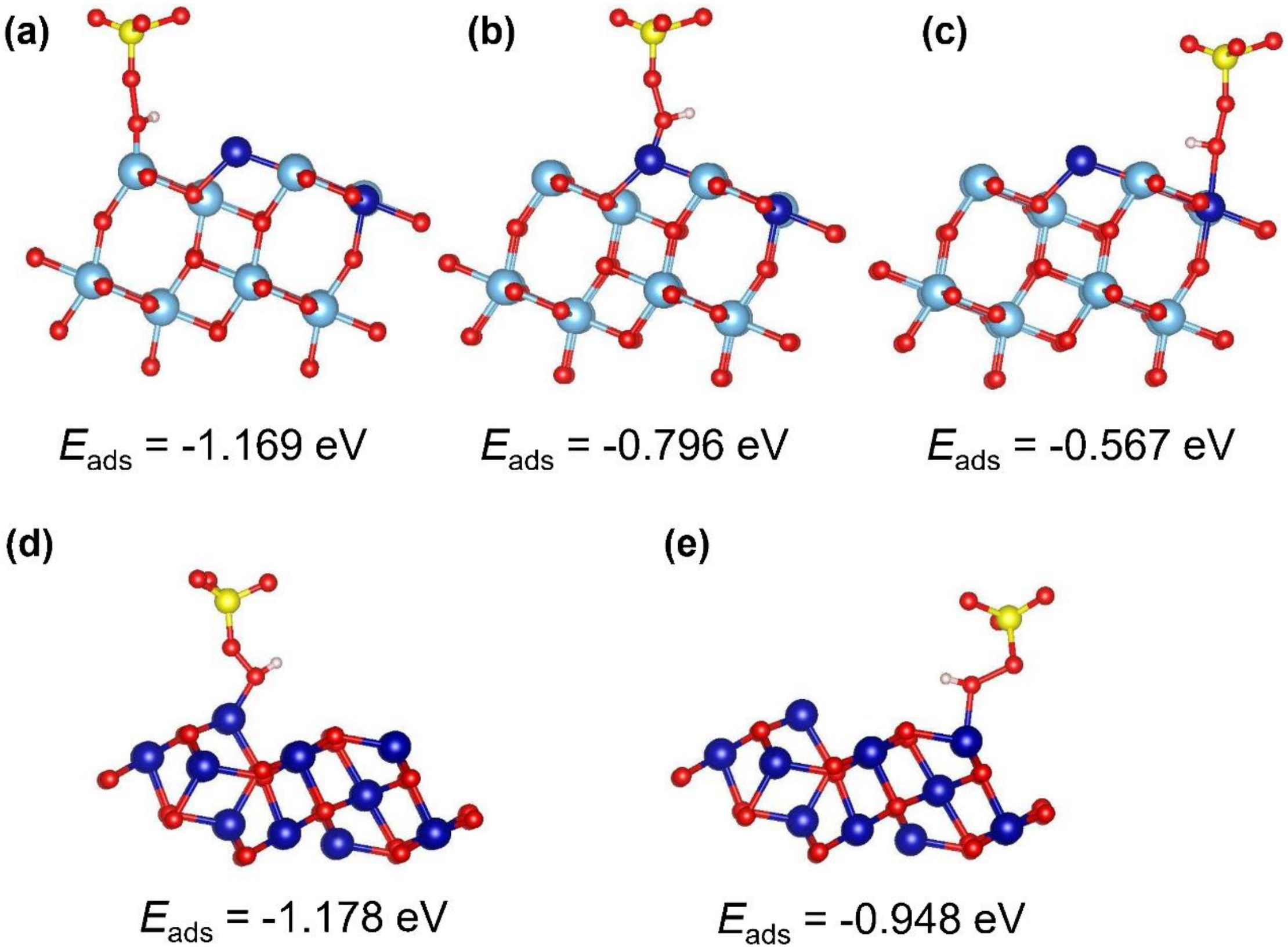

**Fig. 7.** Optimum structure of PMS molecules on Co-$TiO_2$ and $Co_3O_4$ with $Co^{2+}$, $Co^{3+}$, and $Ti^{4+}$ active sites: (a) PMS on Co-$TiO_2$ with $Ti^{4+}$ active site, (b) PMS on Co-$TiO_2$ with $Co^{2+}$ active site, (c) PMS on Co-$TiO_2$ with $Co^{3+}$ active site, (d) PMS on $Co_3O_4$ with $Co^{2+}$ active site, (e) PMS on $Co_3O_4$ with $Co^{3+}$ active site. The baby blue, red, blue, yellow, pink atoms are Ti, O, Co, S, H, respectively.

According to the free radical quenching experiments, $SO_4^{\bullet-}$ radicals played a dominant role in the BAL oxidation within the radical process. Both $SO_4^{\bullet-}$ and $SO_5^{\bullet-}$ were yielded from the cleavage of O-O (HO-$OSO_3$) and O-H (H-$OOSO_3$) bonds on active sites. $SO_5^{\bullet-}$ radical can further generate $SO_4^{\bullet-}$ (eq.5). We examined the formation pathway of $SO_4^{\bullet-}$ and $SO_5^{\bullet-}$ from PMS on Co-$TiO_2$ and $Co_3O_4$ with different metal sites by DFT calculations. The possible reaction pathways are illustrated in **Fig. 8**, with the related reaction energy barriers ($E_b$) for the activation of PMS (Fig. S9) along with the values of $E_b$ and reaction formation energy ($E_f$) (Table S3). For the $SO_4^{\bullet-}$ formation over Co-$TiO_2$, the $E_b$ of the O-O bond cleavage decreased in the order of $Co^{3+}$ > $Ti^{4+}$ > $Co^{2+}$. Generally, the low $E_b$ value manifests a high formation rate. The lowest $E_b$ of 0.200 eV over the $Co^{2+}$ sites showed ultrahigh activity for $SO_4^{\bullet-}$ formation. Meanwhile, the obtained $^{\bullet}OH$ from PMS was bonded to the active sites of Co-$TiO_2$ (Fig. S9). For the $SO_5^{\bullet-}$ formation over Co-$TiO_2$, $Co^{3+}$ sites showed the lowest $E_b$ (0.279 eV) of the H-O bond cleavage from PMS, confirming that $Co^{3+}$ was beneficial for the $SO_5^{\bullet-}$ formation. Accompanying the $SO_5^{\bullet-}$ formation from PMS, the H atom was bonded to the O atom of adjacent active sites. $Co^{2+}$ and $Co^{3+}$ sites in $Co_3O_4$ also displayed the lowest $E_b$ values for the formation of $SO_4^{\bullet-}$ (0.246 eV) and $SO_5^{\bullet-}$ (0.347 eV), respectively. In comparison, the corresponding $E_b$ of $SO_4^{\bullet-}$ and $SO_5^{\bullet-}$ formation over Co-$TiO_2$ is lower than that over $Co_3O_4$, implying that strong interaction between $CoO_x$ clusters and $TiO_2$ support facilitated the formation of $SO_4^{\bullet-}$ and $SO_5^{\bullet-}$. $Ti^{4+}$ in Co-$TiO_2$ had a relatively high $E_b$ of $SO_4^{\bullet-}$ and $SO_5^{\bullet-}$ formation, manifesting that it is difficult to produce radicals from PMS over $Ti^{4+}$ sites. We also calculated the formation energy of $SO_4^{\bullet-}$ and $SO_5^{\bullet-}$. $E_f$ values of $SO_4^{\bullet-}$ over $Co^{2+}$ and of $SO_5^{\bullet-}$ over $Co^{3+}$ in Co-$TiO_2$ and $Co_3O_4$ are below zero. This validated that $Co^{2+}$ and $Co^{3+}$ activated PMS to generate $SO_4^{\bullet-}$ and $SO_5^{\bullet-}$, respectively. These simulation conclusions were very consistent with the experimental results.

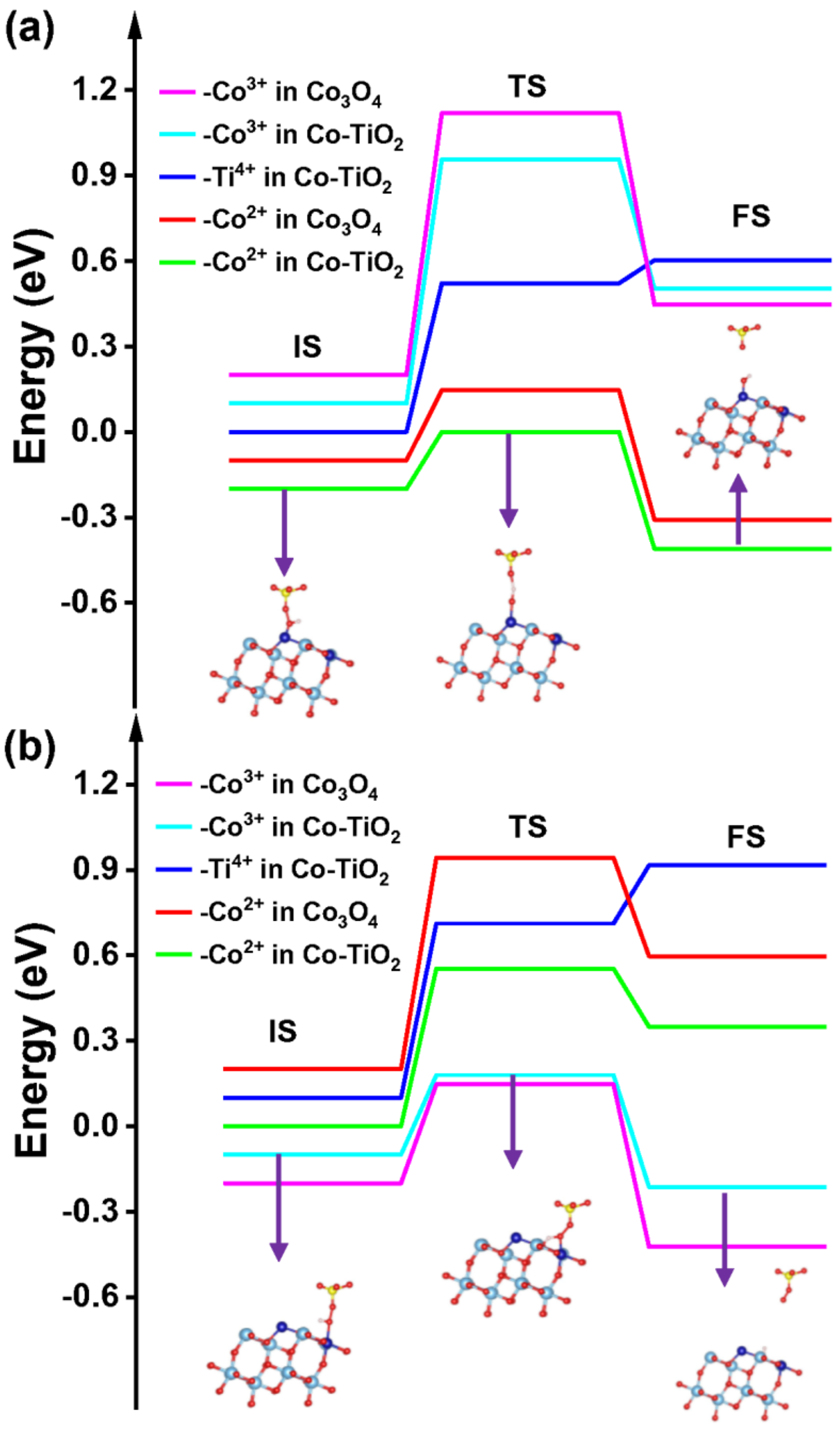

**Fig. 8.** Reaction pathway of the formation of (a) $SO_4^{\bullet-}$ radical and (b) $SO_5^{\bullet-}$ radical through the PMS activation on Co-$TiO_2$ and $Co_3O_4$ with different metal sites, where IS, TS, and FS represent the initial structure, transition structure, and final structure, respectively.

Based on the theoretical and experimental results, the possible mechanism for the BZH formation from BAL over Co-$TiO_2$ is shown in **Fig. 9**. Co atoms displaced the Ti atoms and adsorbed onto the anatase (101) facet as major active sites for BAL selective oxidation via PMS activation. The redox cycle of $Co^{2+}/Co^{3+}$ in Co-$TiO_2$, through electron transfer between Co and PMS, yielded $^{\bullet}OH$, $SO_4^{\bullet-}$, and $SO_5^{\bullet-}$ radicals. Further, the obtained $SO_5^{\bullet-}$ radicals formed the singlet oxygen ($^1O_2$), which along with the free radicals subsequently attacked the BAL molecules to generate BZH.

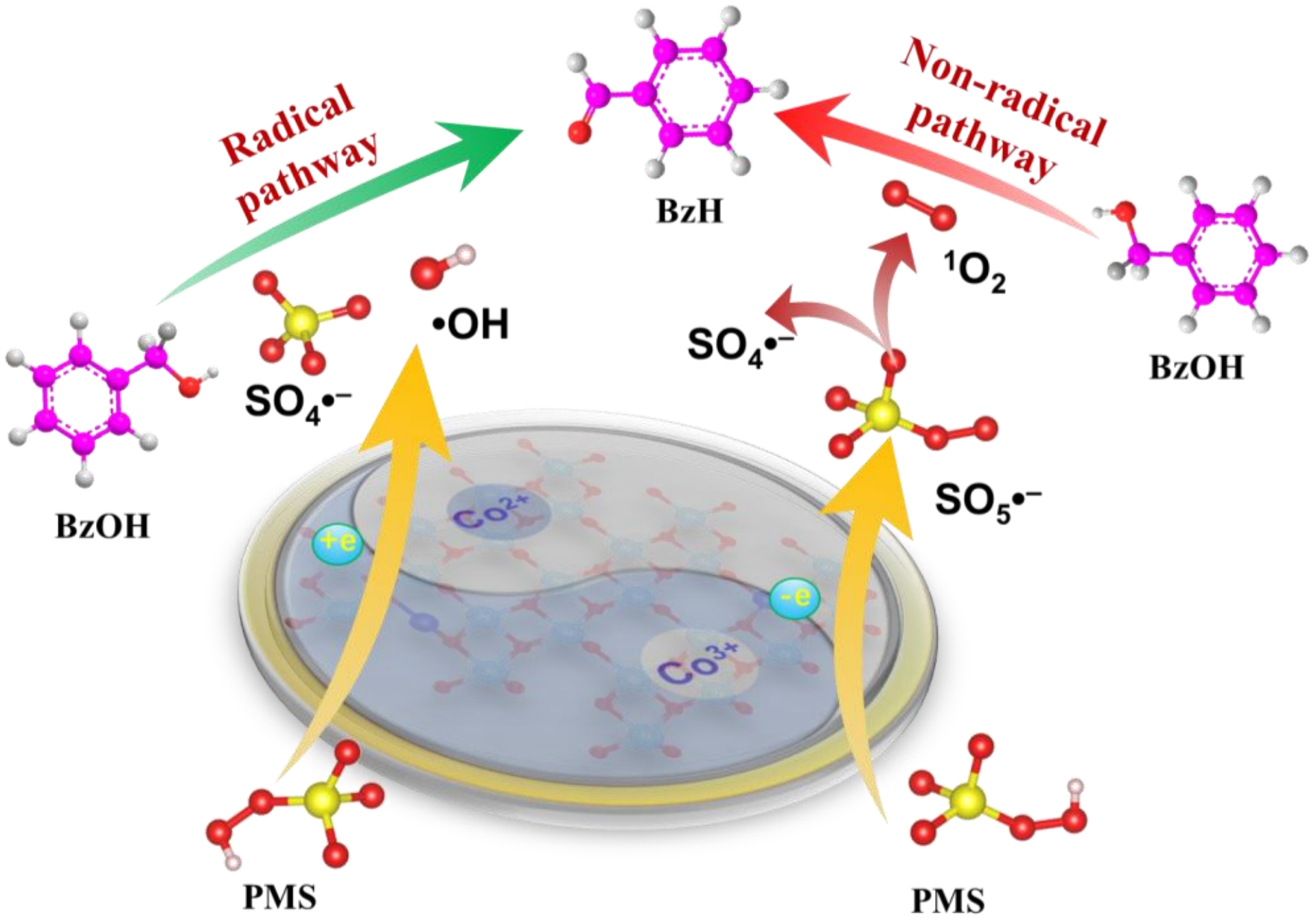

**Fig. 9.** Proposed mechanism of PMS activation and BAL oxidation over Co-$TiO_2$.

## 4. Conclusions

In summary, the subnanometric amorphous $CoO_x$ clusters highly dispersed in anatase $TiO_2$ nanosheets (Co-$TiO_2$) have been successfully fabricated through the green solvent $CO_2$-assisted approach. The as-prepared Co-$TiO_2$ catalyst exhibited superior catalytic performance in the selective conversion of six type of biomass-derived alcohols, whereas high selectivity of aldehydes was achieved. The O vacancies, highly dispersed $CoO_x$ clusters, the strong interaction between Co and $TiO_2$ support were the deciding factors in the catalytic performance. $SO_4^{\bullet-}$ radicals play the dominant role in the directly selective oxidation of BAL and $^1O_2$ participated in the non-radical pathway. DFT calculations further confirmed the experimental findings and showed that the strong interaction between $CoO_x$ clusters and $TiO_2$ support would enhance the formation of $SO_4^{\bullet-}$ and $SO_5^{\bullet-}$. This work offers a sustainable route for the selective oxidation of various alcohols using transition-metal oxide clusters.

**CRediT authorship contribution statement**

**Zhiwei Jiang:** Investigation, Visualization, Writing – original draft; **Zhiyue Zhao**: Software, data curation, formal analysis; **Xin Li**: visualization, validation, review & editing; **Huaiguang Li, Hector F. Garces & Mahmoud Amer**: Writing - review & editing. **Kai Yan**: Supervision, Visualization, Validation, Funding acquisition, Writing –review & editing.

**Declaration of Competing Interest**

The authors declare that they have no known competing financial interests or personal relationships that could have appeared to influence the work reported in this paper.

**Acknowledgements**

This work was supported by National Key R&D Program of China (2020YFC1807600), National Ten Thousand Talent Plan, National Natural Science Foundation of China (21905309 and 22078374), Key-Area Research and Development Program of Guangdong Province (2019B110209003), Guangdong Basic and Applied

Basic Research Foundation (2019B1515120058), the Scientific and Technological Planning Project of Guangzhou, China (202206010145).

**Supplementary material**

Supplementary data to this article can be found on the Elsevier Publications website.